\documentclass{aa}
\usepackage[varg]{txfonts}
\usepackage{natbib}
\usepackage[colorlinks=true,linkcolor=blue,citecolor=blue,filecolor=blue,urlcolor=blue]{hyperref}
\bibpunct{(}{)}{;}{a}{}{,} 
\usepackage{color}
\definecolor{new}{rgb}{0.8,0.2,0.8}
\definecolor{new}{rgb}{0.8,0.2,0.2}

\definecolor{todo}{rgb}{0.8,0.2,0.3}

\definecolor{comm}{rgb}{0,0.7,0}

\definecolor{idea}{rgb}{0,0.5,1}

\definecolor{old}{rgb}{1,0.5,0.5}

\newcommand{\oc}{O\,--\,C}
\newcommand{\kms}{km\,s$^{-1}$}

\begin{document}
\title{OGLE-BLAP-001 and ZGP-BLAP-08: two possible magnetic blue large-amplitude pulsators}
\titlerunning{Two possible magnetic BLAPs}
\author{Andrzej Pigulski\inst{\ref{inst1}} \and Piotr A.~Ko{\l}aczek-Szyma\'nski\inst{\ref{inst1},\ref{inst2},\thanks{This research was supported by the University of Li\`ege under the Special Funds for Research, IPD-STEMA Programme.}} \and Marta \'{S}wi\k{e}ch\inst{\ref{inst1}} \and Piotr {\L}ojko\inst{\ref{inst1}} \and Kacper J.~Kowalski\inst{\ref{inst1}}
}
\institute{Instytut Astronomiczny, Wydzia{\l} Fizyki i Astronomii, Uniwersytet Wroc{\l}awski, Kopernika 11, 51-622 Wroc{\l}aw, Poland \\
\email{Andrzej.Pigulski@uwr.edu.pl}
\label{inst1}
\and
{Space sciences, Technologies and Astrophysics Research (STAR) Institute, Universit\'e de Li\`ege, All\'ee du 6 Ao\^ut
19c, B\^at.~B5c, 4000 Li\`ege, Belgium}\label{inst2}
}
\date{Received date / Accepted date }
\abstract{
Blue large-amplitude pulsators (BLAPs) are a newly discovered group of compact pulsating stars whose origin needs to be explained. Of the existing evolutionary scenarios that could lead to the formation of BLAPs, there are two in which BLAPs are the products of the merger of two stars, either a main sequence star and a helium white dwarf or two low-mass helium white dwarfs. Among over a hundred known BLAPs, three equidistant in frequency modes had been found in one, OGLE-BLAP-001. We show that similar three equidistant in frequency modes exist in yet another BLAP, ZGP-BLAP-08. This perfect separation in frequency is a strong argument to explain the modes in terms of an oblique pulsator model. This model is supported by the character of the changes of the pulsation amplitude and phase with the rotational phase. Consequently, we hypothesize that these two BLAPs are magnetic, as equidistant in frequency pulsation modes should be observed in the presence of a magnetic field whose axis of symmetry does not coincide with the rotation axis. A logical consequence of this hypothesis is to postulate that these two BLAPs could have originated in a merger scenario, just how the origin of magnetic white dwarfs is explained. We also find that period changes in both stars cannot be interpreted by a constant rate of period change and discuss the possible origin of these changes.
}
\maketitle
\authorrunning{A.~Pigulski et al.}
\section{Introduction}
A few years ago, a new group of compact pulsating stars was discovered by \cite{2017NatAs...1E.166P} in the Optical Gra\-vi\-ta\-tio\-nal Lensing Experiment \citep[OGLE,][]{2003AcA....53..291U} data. As these stars are hot and show large amplitudes of pulsations, they have been dubbed blue large-amplitude pulsators (BLAPs). In the Hertzsprung-Russell (H-R) diagram, BLAPs occupy a region between hot massive main-sequence stars and hot subdwarfs. Compared to the latter, however, they have an order of magnitude smaller surface gravities. Their pulsation periods range from 7 to 75 minutes, and their $I$-band peak-to-peak amplitudes are of the order of 0.1\,--\,0.4\,mag and increase for shorter-wavelength passbands. Over a hundred BLAPs and candidate BLAPs are presently known \citep{2023AcA....73..265B,2024arXiv240416089P}.

The light curves of BLAPs are distinctive, with a fast increase in brightness and a much slower decrease, re\-mi\-nis\-cent of some stars pulsating in fundamental radial mode, such as RR\,Lyrae-type stars of the RRab subtype or high-amplitude $\delta$~Scuti-type stars. Indeed, spectroscopic studies and seismic modelling \citep{2017NatAs...1E.166P,2018MNRAS.477L..30R,2020MNRAS.492..232B,2023NatAs...7..223L} have confirmed that the studied BLAPs also pulsate in the fundamental radial modes. Virtually all BLAPs are single-mode pulsators. There are only two exceptions. It is the first discovered BLAP, OGLE-BLAP-001 = OGLE-GB-DSCT-0058, in which the discoverers \citep{2013AcA....63..379P,2015AcA....65...63P} detected two additional small-amplitude sinusoidal components, symmetric with respect to the frequency of the main mode, and recently discovered OGLE-BLAP-030 \citep{2024arXiv240416089P}. In the latter star, these authors found three independent modes, of which two are likely radial. Several low-frequency terms have also been discovered in the photometry of HD\,133729 \citep{2022A+A...663A..62P}, but these terms are most likely {\sl g}-modes originating from the primary component, which is a late B-type star that forms a binary system with the BLAP.

From the beginning of the study of BLAPs, one of the most intriguing challenges has been to explain their origin. To date, several alternative theories for the origin of BLAPs have been proposed. According to these theories, BLAPs could be pre-white dwarfs (pre-WDs) with He cores burning hydrogen in their shells \citep{2017NatAs...1E.166P,2018MNRAS.477L..30R,2018MNRAS.481.3810B,2020MNRAS.492..232B,2018MNRAS.478.3871W}, likely products of a binary stellar evolution \citep{2021MNRAS.507..621B}, stars burning helium in their cores \citep{2017NatAs...1E.166P}, shell-He burning stars \citep{2022A&A...668A.112X}, the surviving components of binary systems in which a white dwarf has exploded as a Type Ia supernova \citep{2020ApJ...903..100M}, or products of a merger of two stars \citep{2023ApJ...959...24Z}. The latter hypothesis seems particularly interesting. \cite{2023ApJ...959...24Z} showed that BLAPs can form as the product of a merger of a He-core white dwarf and a low-mass main-sequence star. In an accompanying paper \citep{2024arXiv241000154K} we show that BLAPs can also be formed as a product of the merger of two extremely low-mass white dwarfs, which was first suggested by \cite{2018arXiv180907451C}.

If some BLAPs are mergers, they can have -- as some other mergers -- strong magnetic fields (Sect.\,\ref{mergers}). Magnetic pulsating stars are known. If magnetic and rotational axes in such stars are inclined, pulsations are usually explained in terms of an oblique pulsator model (Sect.\,\ref{oblique-puls}) because pulsations tend to follow the magnetic axis as the axis of symmetry. Such pulsating stars show distinct pulsation spectra, namely equally spaced terms in the frequency spectrum, similar to those found in OGLE-BLAP-001. In the present paper (Sect.\,\ref{targets}), we report the discovery of this kind of pulsation in another BLAP, ZGP-BLAP-08 \citep{2022MNRAS.511.4971M}. We therefore argue (Sect.\,\ref{sect:discussion}) that these two BLAPs are likely magnetic oblique pulsators. This finding makes the merger origin of some BLAPs highly reliable. The hypothesis that these two stars are magnetic needs to be verified observationally.

\section{Mergers as the origin of strong magnetic fields}\label{mergers}
Stars having measurable magnetic fields are known to have very different masses, spectral types and evolutionary stages \citep[e.g.][]{2015SSRv..191...77F}. In cool stars, magnetic fields originate from surface convection via the hydromagnetic dynamo process \citep{2013SAAS...39.....C}. In hot stars, however, they are usually explained either as fossil fields originating during star formation \citep[e.g.][]{2019EAS....82..345A} or as a consequence of stellar mergers \citep{2009MNRAS.400L..71F,2019Natur.574..211S}. For instance, \cite{2024Sci...384..214F} provided observational evidence that the magnetic field in one of the components of the massive binary HD\,148937 formed via stellar merger. Mergers of young stars can be responsible for the presence of the blue main-sequence band in some young open clusters \citep{2022NatAs...6..480W}.

Magnetic fields in compact stars are also explained by mergers. This, in particular, applies to white dwarfs (WDs), whose magnetic fields originate from an earlier evolution, which includes a merger \citep{2015MNRAS.447.1713B,2018arXiv181006146F} or at least a common envelope phase in the case of cataclysmic binaries \citep{2018MNRAS.481.3604B}. Cool magnetic WDs can also be explained in this way \citep{2020AdSpR..66.1025F}, although a merger with a substellar companion was proposed to interpret their origin. In magnetic WDs formed via mergers, the magnetic fields originate from turbulent accretion on a post-merger star or form during the common envelope phase \citep{2011PNAS..108.3135N,2023LRCA....9....2R}. A massive ($\sim$2\,M$_\odot$) magnetic helium star (HD\,45166) could be formed by a merging of two lower-mass helium stars \citep{2023Sci...381..761S}. This mechanism can explain strong magnetic fields in neutron stars and even the origin of magnetars.

\cite{2023ApJ...959...24Z} have recently shown that BLAPs can also be formed by a merger. In this case, this was a merger of a He-core WD and a low-mass main-sequence companion. As we have already indicated in the Introduction, in the accompanying paper \cite{2024arXiv241000154K} show that BLAPs can also be the products of a different merger, of two extremely low-mass white dwarfs. If any of these two evolutionary channels is realized in nature, magnetic BLAPs may also exist. This possibility was recently confirmed by \cite{2024arXiv240702566P}, who showed that magnetic field can be generated in a merger of two helium WDs. To date, however, no measurements of the magnetic fields of BLAPs have been done.

\section{Oblique pulsator model -- an explanation of pulsations in magnetic stars}\label{oblique-puls}
The best-known pulsating stars explained in terms of an oblique pulsator model are rapidly oscillating Ap (roAp) stars \citep{1982MNRAS.200..807K,1990ARA&A..28..607K}, a subgroup of magnetic chemically peculiar stars. The model assumes that the axis of the symmetry of pulsations is not coaxial with the rotation axis, but follows the magnetic axis, which is inclined with respect to the rotation axis \citep{1982MNRAS.200..807K,1993PASJ...45..617S,2002A&A...391..235B}. This makes the observed amplitudes and phases of the pulsations dependent on the phase of rotation. In the frequency spectra of the photometry of roAp stars, this manifests through the occurrence of equidistant multiplets \citep[e.g.][]{2009MNRAS.396.1189B,2011MNRAS.413.2651B,2021MNRAS.506.1073H,2024MNRAS.527.9548H,2021MNRAS.506.5629S}, usually triplets or quintuplets, separated by the rotational frequency.

While modes excited in roAp stars are mostly dipole or qua\-dru\-po\-le non-radial modes, radially pulsating stars explained by the oblique pulsator model are also known. One of them is $\beta$~Cephei (spectral type B2\,III), the prototype of massive main-sequence stars pulsating in {\sl p}-modes. This star has a measurable longitudinal component of the magnetic field that varies with a rotation period of 12\,d and a peak-to-peak amplitude of about 200\,G \citep{2013A&A...555A..46H}. The pulsations are dominated by a radial mode at frequency 5.25\,d$^{-1}$, but there are other modes in the frequency spectrum of this star which, together with the dominant mode, form an equidistant multiplet \citep{1997A&A...322..493T}. \cite{2000ApJ...531L.143S} interpreted this frequency multiplet in terms of the oblique pulsator model. An oblique rotator model was suggested earlier to explain the variability of the magnetic field in $\beta$~Cep \citep[e.g.][]{1993npsp.conf..186H}. Although \cite{2000ApJ...531L.143S} assumed that the rotation period is 6\,d, half of the true rotation period, such a multiplet can be easily reinterpreted for a rotation period of 12\,d, assuming that some of its components are not detected. 

\section{Target stars and their photometry}\label{targets}
In this section, we present the results of the analysis of the available photometric data for the two BLAPs showing equidistant terms in their frequency spectra. While OGLE-BLAP-001 (Sect.\,\ref{ogle-blap-001}) was known to show equidistant terms, finding such terms in the photometry of ZGP-BLAP-08 (Sect.\,\ref{zgp-blap-08}) is a new discovery.


\subsection{OGLE-BLAP-001}\label{ogle-blap-001}
This star was discovered as variable by \cite{2013AcA....63..379P} during their search for pulsating stars in 21 Galactic disk fields covered by the third phase of the OGLE project, OGLE-III \citep{2003AcA....53..291U}. The star showed a non-sinusoidal variability with a peak-to-peak amplitude of 0.35\,mag in the $I$ band and a short period of 0.0196\,d (28.3\,min). Given the resemblance of its light curve to those of high-amplitude $\delta$~Sct stars, the star was temporarily classified as a $\delta$~Sct star and designated OGLE-GD-DSCT-0058. In addition to the dominant frequency of 50.964\,d$^{-1}$ and its harmonics, \cite{2013AcA....63..379P} found two additional peaks equally separated from the main frequency by $\pm$1.903\,d$^{-1}$. The follow-up low-resolution spectroscopy of this star carried out by \cite{2015AcA....65...63P} showed that the star has an effective temperature of about 33\,000\,K and surface gravity of $\log(g/({\rm cm\,s}^{-2}))=5.3\pm 0.2$\,dex. These properties fitted neither $\delta$\,Sct-type stars nor any other class of known pulsating stars. This prompted \cite{2017NatAs...1E.166P} to search for other similar stars in the OGLE data. This allowed them to define a new class of pulsating compact stars -- BLAPs. As the first discovered member of the class, OGLE-GD-DSCT-0058 has been renamed OGLE-BLAP-001. 
\begin{figure}[!ht]
\includegraphics[width=\columnwidth]{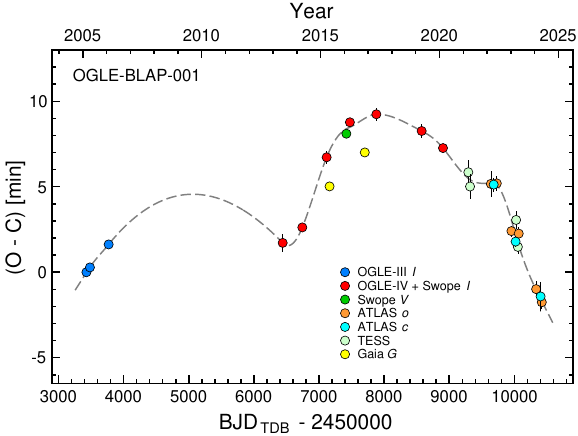}
\caption{{\oc} diagram for the dominant period of OGLE-BLAP-001 plotted as a function of time. The observed times of maximum light (O) and epochs $E$ are given in Table \ref{tab:tmax-ob1}. The predicted times of maximum light (C) were calculated according to Eq.\,(\ref{eph:ob1}). The sources of data are labelled in the plot. The dashed line stands for smoothed changes, which were used to correct the times of observations for period changes, before the final analysis presented in Sect.\,\ref{subs:ob1-final}.}
\label{fig:o-c-o1}
\end{figure}

In the following years, many new members of the class were discovered \citep{2022MNRAS.511.4971M,2022MNRAS.513.2215R,2022A+A...663A..62P,2023NatAs...7..223L,2023AcA....73....1B,2023AcA....73..265B,2024MNRAS.529.1414C,2024arXiv240416089P}, bringing the total number to more than a hundred. None, however, was announced to show the presence of equidistant terms similar to those observed in OGLE-BLAP-001. 

\subsubsection{OGLE and Swope photometry}\label{sect:ob1-ogle}
We have reanalyzed the photometry of OGLE-BLAP-001 published by \cite{2013AcA....63..379P,2017NatAs...1E.166P,2024arXiv240416089P} and made available through the OGLE web page\footnote{\url{http://ogle.astrouw.edu.pl}}. The data set consists of two seasons (2005, 2006) of the OGLE-III $I$-band observations, eight seasons of the OGLE-IV $I$-band observations (2013\,--\,2020) and a few additional nights made with the 1.0-m Swope telescope in 2016 \citep{2017NatAs...1E.166P} in the $V$ and $I$ bands. The analysis showed that the dominant period is slightly variable. Therefore, we split the data into seasons and constructed the {\oc} diagram shown in Fig.\,\ref{fig:o-c-o1}. (Seasons 2017 and 2018 were combined due to the small number of data points.) Before the analysis, the times of observations were converted from HJD to BJD in the standard of Barycentric Dynamical Time, TDB, using the online tool\footnote{\url{https://astroutils.astronomy.osu.edu/time/hjd2bjd.html}} provided by J.~Eastman \citep{2010PASP..122..935E}. The ephemeris we used to construct the {\oc} diagram was the following:
\begin{equation}\label{eph:ob1}
T_{\rm max} = \mbox{BJD}_{\rm TDB}~2453427.40341 + 0.019621473\,\times\,E,
\end{equation}
where $E$ is the number of cycles elapsed from the initial epoch. The times of the maximum light of the dominant term with frequency of 50.96\,d$^{-1}$, derived from seasonal data, are given in Table \ref{tab:tmax-ob1}.

\subsubsection{TESS photometry}\label{subs:ob1-tess}
We also checked if the variability of OGLE-BLAP-001 can be detected in the Transiting Exoplanet Survey Satellite \citep[TESS,][]{2015JATIS...1a4003R} data. The star is TIC\,915565212 in the TESS Input Catalogue, but it was not included as a target for short-cadence photometry with TESS. Therefore, the photometry can be done only with the TESS Full Frame Images (FFIs). 

\begin{figure}[!ht]
\centering
\includegraphics[width=0.48\columnwidth]{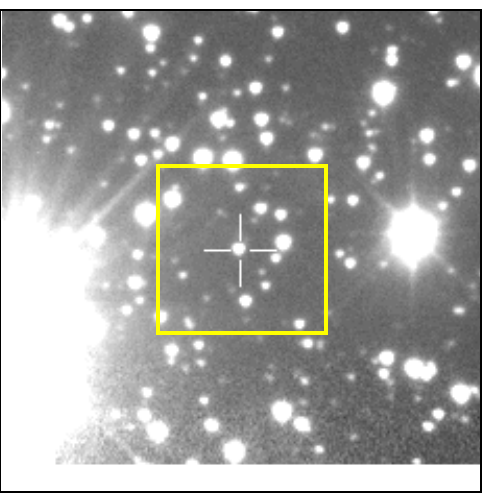}
\includegraphics[width=0.48\columnwidth]{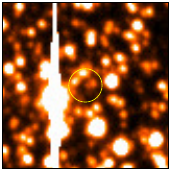}
\caption{Left: $I$-band 1$\arcmin$\,$\times$\,1$\arcmin$ finding chart for OGLE-BLAP-001, centred on the star (white cross), taken from the OGLE web page. The yellow square drawn on it shows the size of a single pixel of the TESS camera (21$\arcsec$\,$\times$\,21$\arcsec$). Right: Part of the ATLAS $o$-band image in the vicinity of OGLE-BLAP-001. The image size is 2.5$\arcmin$\,$\times$\,2.5$\arcmin$. The yellow circle is centred at the target. North is up, East to the left.}
\label{fig:maps-OB1}
\end{figure}

By now, the field of OGLE-BLAP-001 was observed by TESS in six sectors: Sectors 10 and 11 with the FFI cadence of 30\,min, Sectors 36 and 37 with the FFI cadence of 10\,min and Sectors 63 and 64 with the FFI cadence of 200\,s. Two important factors lead to a significant reduction in the observed amplitude of a variable star in the TESS FFI observations. The first factor is the contamination associated with the large angular size of the TESS pixel (21$\arcsec$). It can be seen in Fig.\,\ref{fig:maps-OB1} that there are at least several relatively bright stars falling within a single TESS pixel and many more in its vicinity. The second factor is the total exposure time, which -- if it lasts a large part of the pulsation period -- leads to an amplitude reduction. The observed amplitude ($A_{\rm obs}$) to intrinsic amplitude ($A_{\rm int}$) ratio $f = A_{\rm obs}/A_{\rm int}$ due to this effect amounts to $|{\rm sinc}(\pi t_{\rm puls}/t_{\rm exp})| \leq 1$, where $t_{\rm exp}$ is the exposure (cadence) time and $t_{\rm puls}$ is the period of a given sinusoidal term. Although the light curves of BLAPs are not sinusoidal, the term with the period equal to the pulsation period has the largest amplitude. According to the above relation, the reduction factor $f$ for harmonics is usually smaller than for the main term. In other words, the averaging effect suppresses harmonics stronger than the main term. For OGLE-BLAP-001, with its pulsation period of 28.26\,min, the reduction factor $f$ is equal to 0.058, 0.822 and 0.996 for 30-min, 10-min and 200-s cadence, respectively. This means that the variability of OGLE-BLAP-001 is very likely effectively averaged in the TESS 30-min cadence data.

\begin{figure}[!ht]
\centering
\includegraphics[width=\columnwidth]{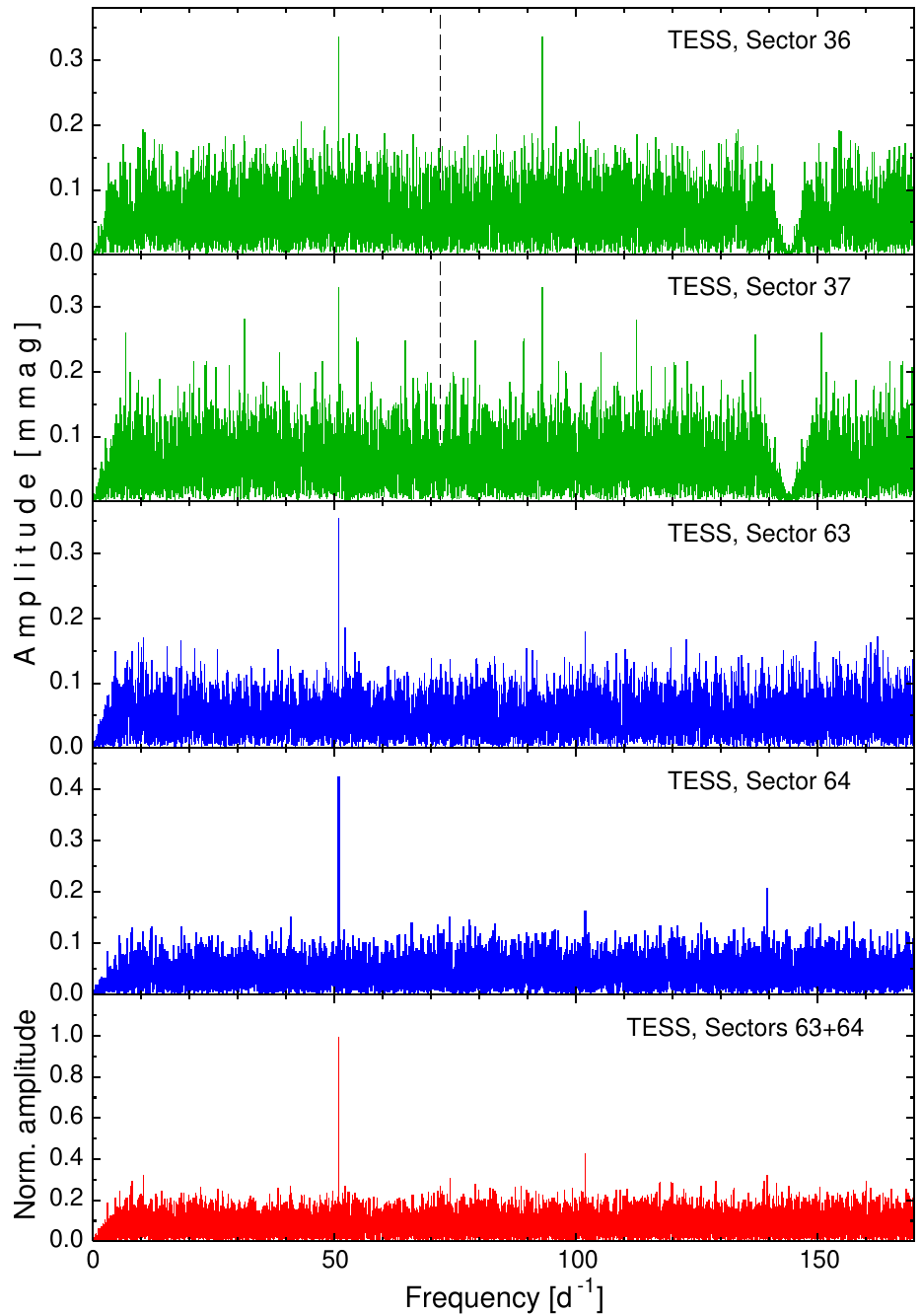}
\caption{Fourier frequency spectra for the pixel photometry of OGLE-BLAP-001 made using TESS FFIs in four sectors (labelled, four upper panels) observed with 10-minute (Sectors 36 and 37) and 200-second (Sectors 63 and 64) cadences and for the combined Sector 63 and 64 data (the lower panel). Drops in signal at low frequencies are due to the strong detrending of the data. The drops are reproduced in the vicinity of the frequency of (10\,min)$^{-1} = 144$\,d$^{-1}$ for the 10-minute cadence data, twice the Nyquist frequency of 72\,d$^{-1}$ marked with the vertical dashed lines.}
\label{fig:ob1-tess-fs}
\end{figure}

Given the brightness of both OGLE-BLAP-001 and ZGP-BLAP-08 ($G\approx17.5$\,mag), their variability is likely at the limits of detection in the TESS data. Nevertheless, we checked this using the Python script written by one of us (PKS). The script uses the {\tt curl} procedure presented in the Mikulski Archive for Space Telescopes (MAST) web page\footnote{\url{https://mast.stsci.edu/tesscut/}}, which was used to download the parts of FFI images surrounding the target \cite[see also][]{2019ascl.soft05007B}. The photometry was obtained by defining pixel(s) in which signal from the target star was expected and pixels in which background was calculated. The total fluxes $F_{\rm tot}$ were obtained by a simple formula:
\begin{equation}
F_{\rm tot} = \sum\limits_{i=1}^N F_i - N\times \mbox{med}\{F_j\},
\end{equation}
where $F_i$ denoted fluxes in $N$ pixels in which we wanted to measure signal and $\mbox{med}\{F_j\}$ stood for the median value of the signal in $j = 1,..., M$ pixels in which background was measured. The errors of the obtained fluxes were calculated using the errors provided for each pixel in FFIs. Then, the fluxes were converted to magnitudes. During analysis, we checked if the pulsation frequency can be detected in the photometry of a single pixel for a 4\,$\times$\,4 pixels window surrounding the position of a given target. Since strong trends occurred in the data, either caused by the variability of the contaminating stars or instrumental effects, we detrended light curves by averaging data in 0.2 and then 0.1-day intervals, interpolating them and then subtracting the smoothed light curve. This efficiently suppressed all low-frequency signals, as can be seen in Fig.\,\ref{fig:ob1-tess-fs}. For Sectors 36, 37 and 63, the variability was detected in a single pixel, and only for Sector 64, in two adjacent pixels. Therefore, the final light curve for Sector 64 was obtained by summing the signal from two pixels. As expected, we did not detect any variability in the 30-minute cadence observations in Sectors 10 and 11 because the pulsation period is very close to the cadence time, which effectively averaged the variability. However, for the remaining four sectors, for which $f$ does not differ much from 1, we were able to detect the frequency of the main term and even its harmonic in the shortest-cadence Sector 63 and 64 data (Fig.\,\ref{fig:ob1-tess-fs}). This is well seen in the combined Sector 63 and 64 data, normalized according to the derived amplitudes (bottom panel of Fig.\,\ref{fig:ob1-tess-fs}). The successful detection allowed us to determine the times of maximum light separately for the four sectors. The TESS data significantly extend the time range for which we can track changes in the pulsation period of OGLE-BLAP-001 in the {\oc} diagram (Fig.\,\ref{fig:o-c-o1}). As can be seen in Fig.\,\ref{fig:o-c-o1}, the long-term period changes in OGLE-BLAP-001 are even more complicated than deduced from the OGLE and Swope data.  

\subsubsection{ATLAS photometry}\label{subs:ob1-atlas}
The Asteroid Terrestrial-impact Last Alert System (ATLAS) \citep{2018AJ....156..241H,2018PASP..130f4505T} is an early warning system focused on potentially hazardous asteroids. It covers the whole sky with four 50-cm telescopes located in observatories in Hawaii (two telescopes), Chile and South Africa. The observations we used in our analysis were obtained by means of the ATLAS Forced Photometry Server\footnote{\url{https://fallingstar-data.com/forcedphot/}} with differential images \citep{2020PASP..132h5002S}. We chose this kind of photometry because ATLAS images for OGLE-BLAP-001 are crowded due to the 1.86$\arcsec$ per pixel scale of the ATLAS images \citep{2018PASP..130f4505T} (Fig.\,\ref{fig:maps-OB1}).

The observations we used covered a time interval between 1 January 2022 and 7 June 2024. They were obtained in two ATLAS passbands, $c$ (cyan) and $o$ (orange) \citep{2018PASP..130f4505T}. This resulted in 240 $c$-band and 805 $o$-band observations. The ATLAS photometry is given as differential fluxes expressed in $\mu$Jy. At the beginning of the analysis, we removed observations with errors larger than 100\,$\mu$Jy. Then, observations were converted to magnitudes assuming an arbitrary value of the (unknown) total flux from OGLE-BLAP-001. Since times provided with the ATLAS photometry were expressed in Modified Julian Days (MJD) and were given for the beginning of an exposure, we converted them to BJD$_{\rm TDB}$, adding also half of the 30-second exposure time. Subsequently, we searched for periodic signals in the data, applying also some weak detrending and removing outliers via 3-$\sigma$ clipping. The final light curves consisted of 221 $c$-band and 716 $o$-band data points. We detected $f_1$ and 2$f_1$ in both data sets. Then, the $c$ and $o$-band data were split into three seasons (2022, 2023 and 2024), and the more abundant $o$-band data were additionally split into two parts in each season. These seasonal data were used to derive nine times of maximum light, which are given in Table \ref{tab:tmax-ob1}. The {\oc} values resulting from these times of maximum light are shown in Fig.\,\ref{fig:o-c-o1}. As can be seen, the ATLAS data fit pretty well the {\oc} changes obtained with the TESS data.

\subsubsection{{\it Gaia} DR3 photometry}\label{subs:ob1-gaia}
In {\it Gaia} Data Release 3 (GDR3) data the star has a number 5254042907771535616. It was recognized as a large-amplitude variable from {\it Gaia} data \citep{2021A&A...648A..44M} and classified either as DSCT/GDOR/ SXPHE or a short-timescale variable (S) in {\it Gaia} catalogues \citep{2023A&A...674A..13E}. The GDR3 data set consists of 46 $G$-band data points and 36 data points in each of the $G_{\rm BP}$ and $G_{\rm RP}$ passbands gathered between July 2014 and May 2017. Since {\it Gaia} DR3 times are given in BJD, Barycentric Coordinate Time (TCB) standard, we transferred the BJD$_{\rm TCB}$ to BJD$_{\rm TDB}$ according to the relation provided in the International Astronomical Union (IAU) 2006 Resolution B3\footnote{\url{https://www.iau.org/static/resolutions/IAU2006\_Resol3.pdf}}. For the {\it Gaia} DR3 observations, the difference between the two times amounts to 18\,--\,20\,s.  In addition, we split {\it Gaia} $G$ data into two parts, which resulted in two times of maximum light given in Table \ref{tab:tmax-ob1}. As can be seen in Fig.\,\ref{fig:o-c-o1}, the {\it Gaia} points lag behind those of OGLE by $\sim$2 min. We cannot explain this discrepancy. It can be related to the differences in passbands of the data we used or the scarcity of the {\it Gaia} data.

\subsubsection{Analysis of the combined photometry}\label{subs:ob1-final}
It is clear from Fig.\,\ref{fig:o-c-o1} that the dominant pulsation period in OGLE-BLAP-001 changes. During $\sim$19 years covered by observations, the full range of these changes amounts to about 12 minutes, that is, about 40\% of the pulsation period. We discuss the possible reasons for the period changes in Sect.\,\ref{sect:discussion}. Because we are interested in the detection of low-amplitude terms other than the dominant pulsation, we decided to combine the available photometry to increase their detectability. In general, such a procedure should lead to an increase in detectability provided that all periodic terms change periods in a similar way and that the shapes of the light curve in different passbands are not very different. This is not known a priori, but the final analysis should show whether these assumptions are valid. Before the final analysis, we carried out the following steps: (i) The photometric subsets discussed in Sect.\,\ref{sect:ob1-ogle}\,--\,\ref{subs:ob1-gaia} were corrected for period changes. For this purpose, we have calculated averages of nearby {\oc} values and interpolated between them using Akima interpolation \citep{1991ACMT...17..367A,1991ACMT...17..341A}. This resulted in the curve shown with the dashed line in Fig.\,\ref{fig:o-c-o1}. This smooth curve was used to correct all times of observations for period changes.\footnote{The curve does not represent any specific model of period changes and therefore does not address the physical reason for these changes. It only serves as a smooth approximation of the period changes.} (ii) The corrected photometric subsets were analyzed separately to derive mean magnitudes, amplitudes of the dominant term and residual standard deviation. (iii) For each subset, we first subtracted the mean magnitudes. Then, the data were normalized to the amplitude of the main term with the frequency of 50.95\,d$^{-1}$ and finally the errors equal to the residual standard deviation normalized to the same amplitude were assigned to all observations. The latter step was done because the original errors in some data sets were found to be underestimated. (iv) All suitable subsets were combined. 

\begin{figure*}
\centering
\includegraphics[width=0.9\textwidth]{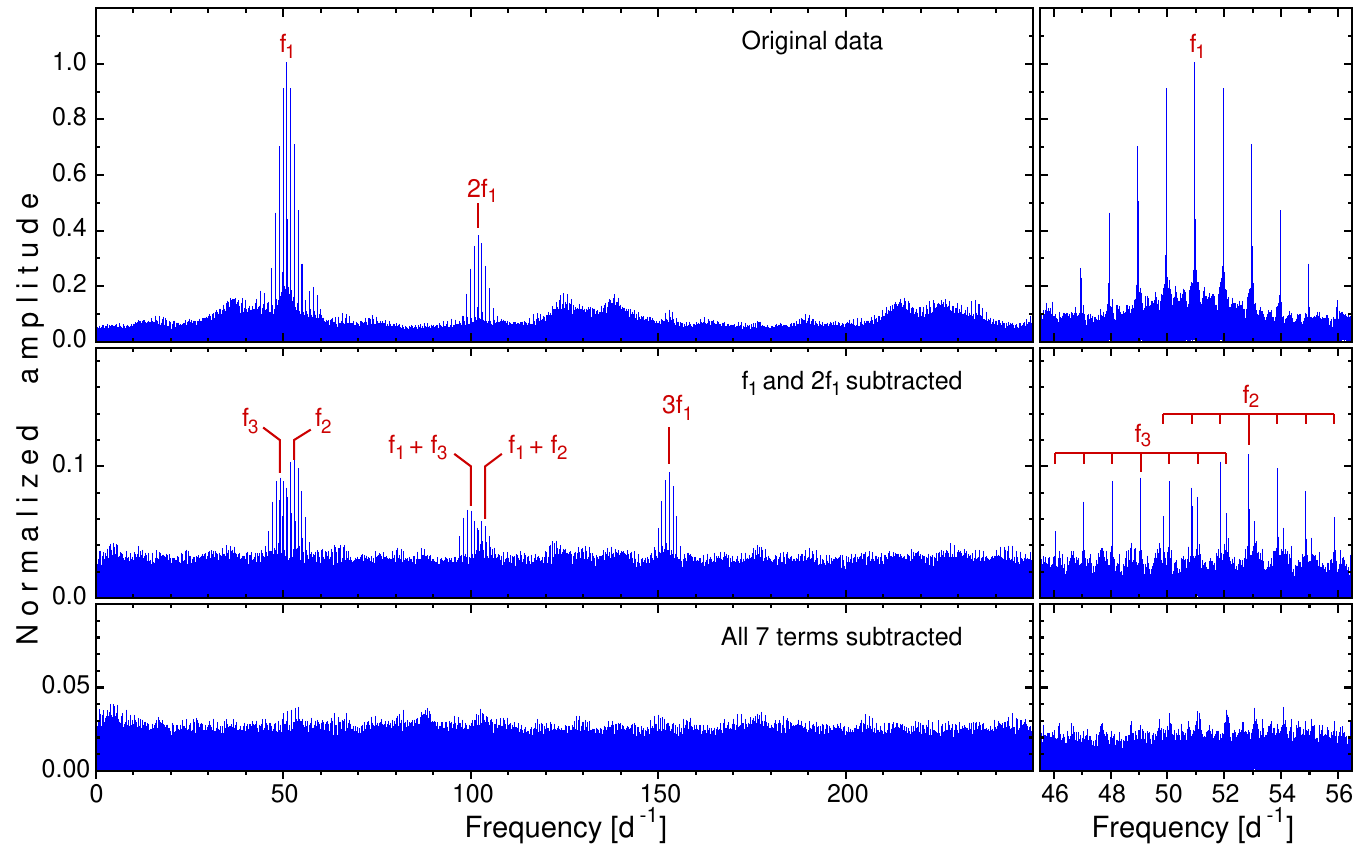}
\caption{Fourier frequency spectra of the combined data of OGLE-BLAP-001 at three steps of prewhitening: for the original data (top panels), after subtracting the dominant term $f_1$ and its strongest harmonic, 2$f_1$ (middle panels), and after subtracting all seven detected terms (bottom panels). The left panels show spectra in the range of 0\,--\,250\,d$^{-1}$, and the right panels, zoomed parts of the frequency spectra close to the dominant term $f_1$. The middle right panel shows the location of $f_2$ and $f_3$ and their nearest daily aliases, marked with downward tics.}
\label{fig:ob1-fspectra}
\end{figure*}

We finally decided not to include {\it Gaia} and TESS data into the final data set. While the {\it Gaia} data are very sparse, the TESS data are affected by strong contamination. The final combined light curve of OGLE-BLAP-001 included OGLE-III $I$, OGLE-IV $I$, Swope $VI$ and ATLAS $co$ observations and consisted of 2886 data points. This combined light curve was the subject of a standard prewhitening procedure in which a variability model consisting of $L$ sinusoidal terms
\begin{equation}\label{eq:sin}
\sum\limits_{i=1}^L A_i\sin(2\pi f_i t + \phi_i)
\end{equation}
was fitted to the data. In the consecutive steps, the subsequent frequencies were derived from the Fourier spectra of the residuals of the previous fit. One term was added to the variability model in each iteration. In Eq.\,(\ref{eq:sin}), $A_i$, $f_i$, and $\phi_i$ stand for semi-amplitude, frequency and phase of the $i$-th term, respectively. The results of the least-squares fit of this variability model to the combined data of OGLE-BLAP-001 are given in Table \ref{tab:ob1-final}. The signal-to-noise ratio (S/N) values, (S/N)$_i = {\rm S}_i/S_{\rm noise}$, were calculated assuming $S_i=A_i$, while the noise $S_{\rm noise}=0.00877$ was calculated as the average signal in the Fourier spectrum of the final residuals in the range 0\,--\,250\,d$^{-1}$. Fourier spectra of the combined data at three steps of prewhitening are shown in Fig.\,\ref{fig:ob1-fspectra}.

We have detected seven terms in the combined data, including three independent terms with the same frequencies as found by \cite{2013AcA....63..379P}. The splittings are equal to $\Delta_1=f_2-f_1 = 1.902964(11)$\,d$^{-1}$ and $\Delta_2=f_1-f_3= 1.902928(13)$\,d$^{-1}$, the difference between the two splittings is equal to 0.000036(17)\,d$^{-1}$. Therefore, we can conclude that within 2-$\sigma$, the two splittings are the same, and the frequencies of the three independent terms in OGLE-BLAP-001 are equidistant. 
\begin{figure}[!ht]
\centering
\includegraphics[width=\columnwidth]{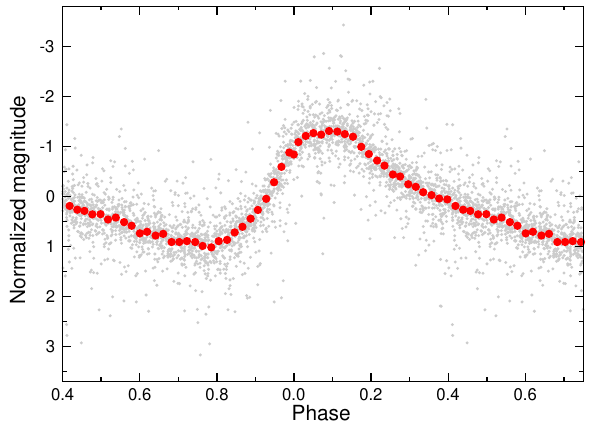}
\caption{Phased light curve for the dominant pulsation mode in OGLE-BLAP-001. The data were corrected for period changes, freed from the contribution of the terms other than $f_1$ and its harmonics, and then combined and phased with the period of $P_1 = 1/f_1 = 0.0196214703$\,d. Red dots stand for average values calculated in 0.02 phase intervals. Phase 0.0 corresponds to BJD$_{\rm TDB}$\,2457000.0.}
\label{fig:ob1-phase}
\end{figure}

\begin{table}
\caption{Parameters of the fit of the variability model to the OGLE-BLAP-001 data.}
\label{tab:ob1-final}
\centering
\small
\begin{tabular}{cr@{.}lr@{.}lr@{.}lr@{.}l}
\hline\hline\noalign{\smallskip}
\multicolumn{1}{c}{Term} & \multicolumn{2}{c}{$f_i$} & \multicolumn{2}{c}{$A_i$} & \multicolumn{2}{c}{$\phi_i$} & \multicolumn{2}{c}{${\rm (S/N)}_i$}\\
 & \multicolumn{2}{c}{[d$^{-1}$]} & \multicolumn{2}{c}{} & \multicolumn{2}{c}{[rad]} & \multicolumn{2}{c}{}\\
\noalign{\smallskip}\hline\noalign{\smallskip}
$f_1$ & 50&9645803(10) & 1&000(12) & 3&770(12) & 114&0 \\
2$f_1$ & 101&9291605 & 0&376(12) & 3&853(32) & 42&9 \\
$f_2$ & 52&867544(11) & 0&112(12) & 6&25(11) & 12&8 \\
$f_3$ & 49&061653(13) & 0&093(12) & 2&04(13) & 10&6 \\
3$f_1$ & 152&8937408 & 0&098(12) & 3&90(12) & 11&2 \\
$f_1+f_3$ & 100&026233 & 0&066(12) & 1&72(18) & 7&5 \\
$f_1+f_2$ & 103&832124 & 0&053(12) & 0&64(23) & 6&0 \\
\noalign{\smallskip}\hline 
\end{tabular}
\tablefoot{The data used in the fit are the combined OGLE, Swope and ATLAS data, corrected for period changes and normalized to the amplitude of the $f_1$ term. Phases are given for epoch BJD$_{\rm TDB}$\,2457000.0.}
\end{table}


\subsection{ZGP-BLAP-08 (OGLE-BLAP-061)}\label{zgp-blap-08}
The star, named ZTF\,J191308.77+120451.6, was discovered as a variable by \cite{2020MNRAS.499.5782O} in the Zwicky Transient Facility (ZTF) Data Release 1 (DR1) data \citep{2019PASP..131a8002B,2019PASP..131a8003M}, but the authors did not make a classification. Using ZTF DR2 data, \cite{2020ApJS..249...18C} also found this star variable, derived a period of 0.02501\,d (36.0\,min) and included it in the table of suspected variable stars. \cite{2022MNRAS.511.4971M} made a careful selection of candidates for BLAPs using {\it Gaia} DR2 photometry \citep{2018A&A...616A...1G} and Pan-STARRS\,1 3$\pi$ Survey \citep{2020ApJS..251....6M} 3D dust maps of \cite{2019ApJ...887...93G}. Then, using ZTF DR3 time-series photometry they searched for BLAPs and found the star a high-confidence BLAP candidate with a period of 35.137\,min. It was named ZGP-BLAP-08. Recently, \cite{2023AcA....73....1B} found the variability of this star with the same period in the OGLE-IV Galactic disk data and also classified it as a BLAP, assigning it the name OGLE-BLAP-061.
\begin{figure}
\includegraphics[width=\columnwidth]{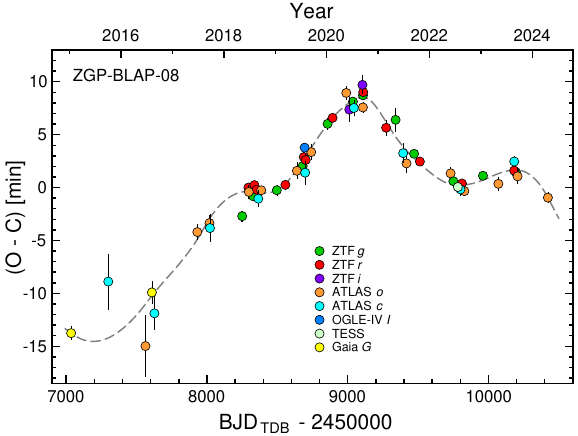}
\caption{{\oc} diagram for the dominant period of ZGP-BLAP-08 plotted as a function of time. The observed times of maximum light (O) and epochs $E$ are given in Table \ref{tab:tmax-zb8}. The predicted times of maximum light (C) were calculated according to Eq.\,(\ref{eph:zb8}). The sources of data are labelled in the plot. The dashed line stands for smoothed changes, which were used to correct the times of observations for period changes, before the final analysis presented in Sect.\,\ref{subs:zb8-final}.}
\label{fig:o-c-zb8}
\end{figure}

\subsubsection{ZTF photometry}
We started our analysis of the photometry of ZGP-BLAP-08 from the ZTF data because the star was discovered as a BLAP using these data. At the time of writing, ZTF DR21 data were available, covering the time interval of almost six years between 27 March 2018 and 28 February 2024. In total, 1176, 2299 and 155 data points were available in $g$ (three subsets), $r$ (three subsets), and $i$ (one subset) passbands, respectively. At the first step, the HJD times provided with the data were converted to BJD$_{\rm TDB}$ times.

We first analyzed data from each of the seven subsets independently. During analysis, we removed some outliers with 3-$\sigma$ clipping and removed some small seasonal trends. Then, we combined all $g$-band data accounting for the differences in mean magnitudes and did the same with the three $r$-band subsets. These $g$ and $r$-band data were then divided into 12 subsets, each containing a similar number of data points, to which a model consisting of the main term with frequency of about 40.98\,d$^{-1}$  and its harmonics was fitted. This resulted in 24 times of maximum light, which are presented in Table \ref{tab:tmax-zb8}. Since ZTF $i$-band data were made only in one season (2020), we split the data into two parts, which resulted in two additional times of maximum light (Table \ref{tab:tmax-zb8}). The corresponding {\oc} diagram was then calculated using the following ephemeris:
\begin{equation}\label{eph:zb8}
T_{\rm max} = \mbox{BJD}_{\rm TDB}~2458294.62233 + 0.024400817\,\times\,E.
\end{equation}
The diagram is shown in Fig.\,\ref{fig:o-c-zb8}. It can be seen that the period of the dominant pulsation term is variable. The range of the {\oc} values from the ZTF data amounts to about nine minutes, that is, about 25\% of the pulsation period. That the pulsation period of this star is changing was found during the preliminary analysis of the ZTF data, as a strong signal near the frequency of the subtracted dominant term remained.

\subsubsection{ATLAS photometry}
\begin{figure}[!ht]
\centering
\includegraphics[width=0.48\columnwidth]{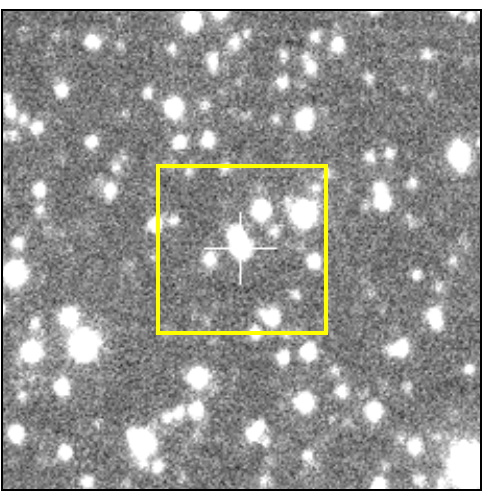}
\includegraphics[width=0.48\columnwidth]{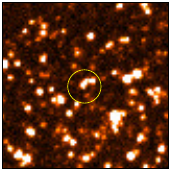}
\caption{Left: $I$-band 1$\arcmin$\,$\times$\,1$\arcmin$ finding chart for ZGP-BLAP-008, centred on the star (white cross), taken from the OGLE web page. The yellow square drawn on it shows the size of a single pixel of the TESS camera (21$\arcsec$\,$\times$\,21$\arcsec$). Right: Part of the ATLAS $o$-band image in the vicinity of ZGP-BLAP-008. The image size is 2.5$\arcmin$\,$\times$\,2.5$\arcmin$. The yellow circle is centred at the target. North is up, East to the left.}
\label{fig:maps-ZB8}
\end{figure}

The ATLAS differential photometry of ZGP-BLAP-08 was processed in the same way as for OGLE-BLAP-001 (Sect.\,\ref{subs:ob1-atlas}). Observations we used covered time interval between 26 July 2015 and 3 June 2024 and consisted of 3420 $o$-band and 896 $c$-band data points. The star is located in a less crowded field than OGLE-BLAP-001 (Fig.\,\ref{fig:maps-ZB8}). We carried out a preliminary analysis rejecting outliers and removing small trends in the data. This resulted in 3192 and 850 data points in the final $o$ and $c$-band data, respectively. The data were then split into seasons. Some scarcely-populated seasons were combined, while densely-populated $o$-band seasons were split into two parts. This resulted in fifteen $o$-band and nine $c$-band times of maximum light (Table \ref{tab:tmax-zb8}). The resulting {\oc} values are presented in Fig.\,\ref{fig:o-c-zb8}. It can be seen that ATLAS data extended the coverage in the {\oc} diagram, although 2015 and 2016 data are scarce, which resulted in relatively large uncertainties of the corresponding {\oc} values.

\subsubsection{OGLE photometry}
ZGP-BLAP-08 was found as a BLAP also in OGLE data by \cite{2023AcA....73....1B}, who named the star OGLE-BLAP-061. The OGLE-IV data are scarce and consist of 69 $I$-band data points, three in 2015, eight in 2018 and 58 in 2019. Therefore, only 2019 data were used to derive a single time of maximum light (Table \ref{tab:tmax-zb8}, Fig.\,\ref{fig:o-c-zb8}).

\subsubsection{TESS photometry}
ZGP-BLAP-08 (TIC\,1795171622) was observed by TESS in Sector 54 with 200-second cadence, but similar to OGLE-BLAP-001, the star was not scheduled for short-cadence observations. Therefore, we proceeded in the same way as in the case of OGLE-BLAP-001 (Sect.\,\ref{subs:ob1-tess}), extracting photometry from the TESS FFIs. Since the data cover about a month, a single time of maximum light was derived from the TESS data, which revealed clearly the main term, its harmonic, and even two side terms; see Sect.\,\ref{subs:zb8-final}. The corresponding value of the {\oc} (Table \ref{tab:tmax-zb8}, Fig.\,\ref{fig:o-c-zb8}) fits well with the ZTF and the ATLAS data.
\begin{figure*}
\centering
\includegraphics[width=0.9\textwidth]{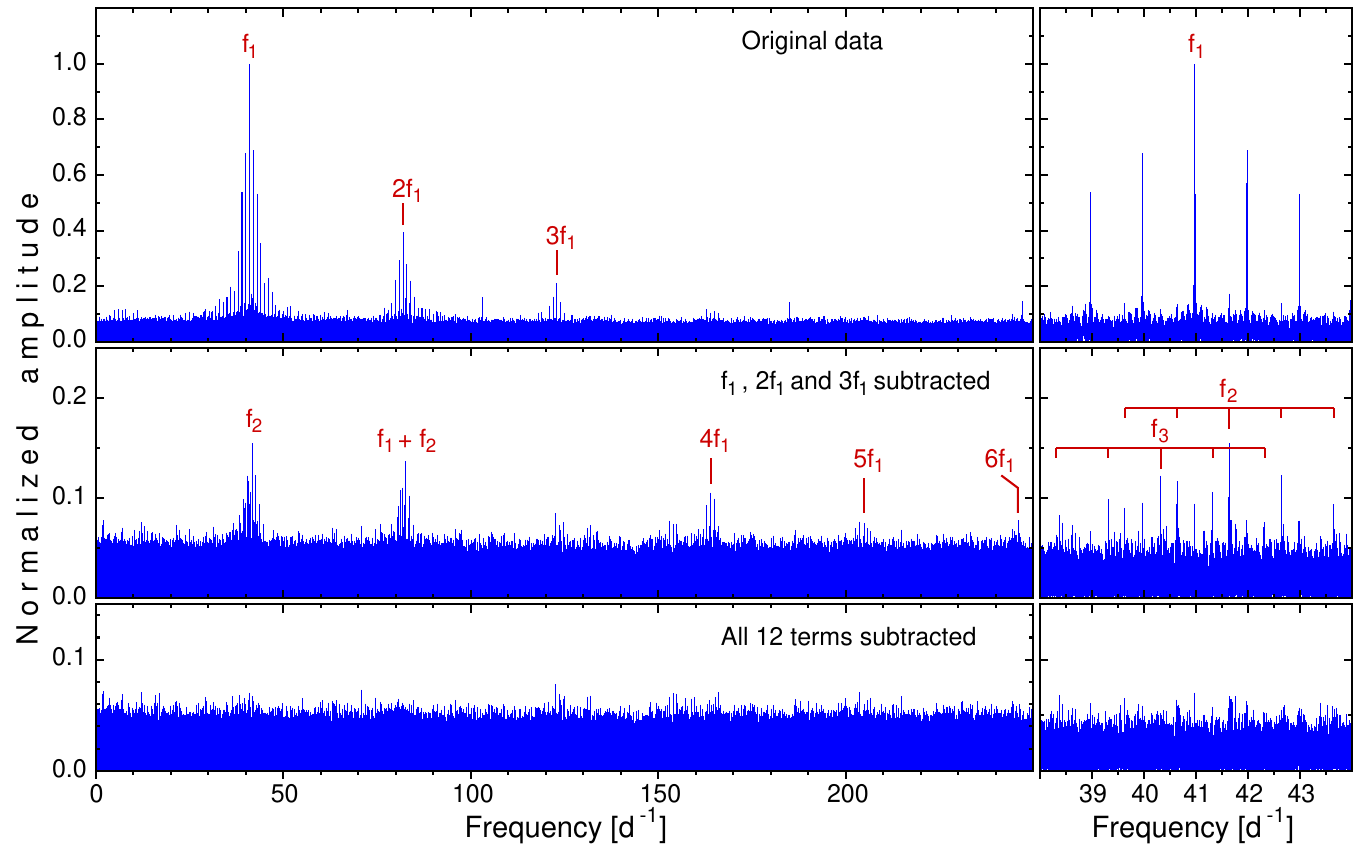}
\caption{Fourier frequency spectra of ZGP-BLAP-08 combined data at three steps of prewhitening: for the original data (top panels), after subtracting the dominant term $f_1$ and its two strongest harmonics (middle panels), and after subtracting all 12 terms (bottom panels). The left panels show spectra in the range of 0\,--\,250\,d$^{-1}$, and the right panels, zoomed parts of the frequency spectra close to the dominant term $f_1$. The middle right panel shows the location of $f_2$ and $f_3$ and their nearest daily aliases, marked with tics.}
\label{fig:zb8-fspectra}
\end{figure*}

\subsubsection{{\it Gaia} DR3 photometry}
The star is GDR3 4313182256636435712. It was recognized as a large-amplitude variable from the {\it Gaia} data \citep{2021A&A...648A..44M} and classified as a short-timescale variable (S) in the {\it Gaia} catalogue \citep{2023A&A...674A..13E}. The GDR3 data set consists of 59 $G$-band data points and 57 data points in each of the $G_{\rm BP}$ and $G_{\rm RP}$ passbands gathered between July 2014 and May 2017. The {\it Gaia} DR3 data are relatively scarce, but because they cover nearly three years, we split them into two parts and derived two times of maximum light which are given in Table \ref{tab:tmax-zb8} and depicted in Fig.\,\ref{fig:o-c-zb8}. Only the least scattered $G$-band data, after removing three data points with the largest errors, were used.
\begin{table}
\caption{Parameters of the fit of the variability model to the ZGP-BLAP-08 data.}
\label{tab:zb8-combined}
\centering
\small
\begin{tabular}{cr@{.}lr@{.}lr@{.}lr@{.}l}
\hline\hline\noalign{\smallskip}
\multicolumn{1}{c}{Term} & \multicolumn{2}{c}{$f_i$} & \multicolumn{2}{c}{$A_i$} & \multicolumn{2}{c}{$\phi_i$} & \multicolumn{2}{c}{${\rm (S/N)}_i$}\\
 & \multicolumn{2}{c}{[d$^{-1}$]} & \multicolumn{2}{c}{} & \multicolumn{2}{c}{[rad]} & \multicolumn{2}{c}{}\\
\noalign{\smallskip}\hline\noalign{\smallskip}
$f_1$ & 40&9822311(26) & 0&997(17) & 0&941(18) & 58&5 \\
2$f_1$ & 81&9644622 & 0&385(17) & 4&47(5) & 22&6 \\
3$f_1$ & 122&9466933 & 0&190(17) & 1&75(9) & 11&2 \\
$f_2$ & 41&639638(21) & 0&134(17) & 3&76(13) & 7&9 \\
$f_1+f_2$ & 82&621869 & 0&131(17) & 0&34(13) & 7&7 \\
$f_3$ & 40&324757(24) & 0&118(17) & 3&90(15) & 6&9 \\
$f_1+f_3$ & 81&306988 & 0&108(17) & 0&96(16) & 6&3 \\
4$f_1$ & 163&9289244 & 0&103(17) & 5&16(17) & 6&0 \\
5$f_1$ & 204&9111555 & 0&077(17) & 1&94(22) & 4&5 \\
6$f_1$ & 245&8933866 & 0&076(17) & 4&63(22) & 4&5 \\
\noalign{\smallskip}\hline\noalign{\smallskip}
$f_2^\prime$ & 41&64102(4) & 0&101(17) & 5&05(18) & 5&9 \\
$f_1^\prime$ & 40&97885(5) & 0&086(17) & 0&66(20) & 5&1 \\
\noalign{\smallskip}\hline 
\end{tabular}
\tablefoot{The data used in the fit are the combined data, corrected for period changes and normalized to the amplitude of the dominant term. Phases are given for epoch BJD$_{\rm TDB}$\,2459200.0.}
\end{table}

\subsubsection{Analysis of the combined photometry}\label{subs:zb8-final}
Figure \ref{fig:o-c-zb8} clearly shows that the pulsation period of ZGP-BLAP-08 changes significantly during these nine years covered by observations. Furthermore, we found that in addition to the main term with a frequency of 40.98\,d$^{-1}$ and its harmonics, there are also other low-amplitude terms. Therefore, we proceeded in the same way as for OGLE-BLAP-001 (Sect.\,\ref{subs:ob1-final}). First, we accounted for period changes by correcting observing times according to the smoothed relation shown with the dashed line in Fig.\,\ref{fig:o-c-zb8}, then subtracted mean magnitudes for each data set, normalized each data set with the amplitude of the dominant term, and finally combined all data. The final data set consisted of 10\,570 data points. This data set was the subject of the final analysis. Since original uncertainties were not always reliable, we assigned the same uncertainty for all data points of a given data set, based on the residual standard deviation of the preliminary analysis of each data set separately, scaled with the amplitude of the dominant term. 

\begin{figure}
\includegraphics[width=\columnwidth]{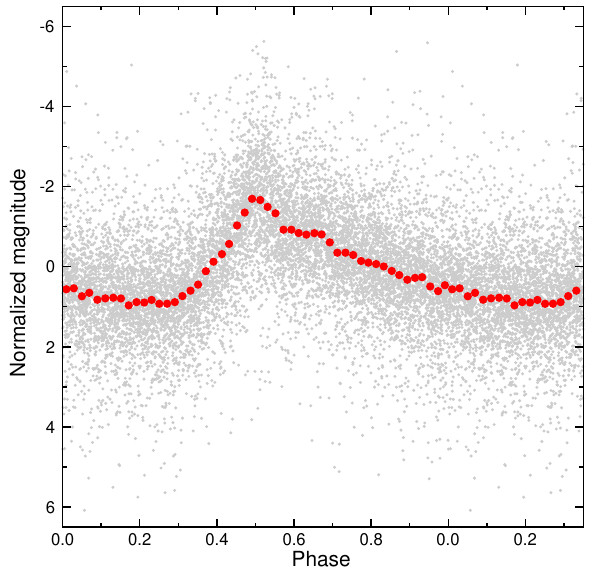}
\caption{Light curve of the combined data of ZGP-BLAP-08 corrected for period changes, freed from the contribution of terms other than $f_1$ and its harmonics and phased with the period of $P_1 = 1/f_1 = 0.0244008189$\,d. Red dots stand for average values calculated in 0.02 phase intervals. Phase 0.0 corresponds to BJD$_{\rm TDB}$ 2459200.0.}
\label{fig:zb8-phased}
\end{figure}

The results of the final analysis are shown in Fig.\,\ref{fig:zb8-fspectra} and Table \ref{tab:zb8-combined}. The values of S/N were calculated in the same way as for OGLE-BLAP-001. For ZGP-BLAP-08 combined data, $S_{\rm noise}$ (noise) was equal to 0.0171. In total, we found 12 terms, including the dominant term with frequency $f_1$, and two other independent frequencies, $f_2$ and $f_3$\footnote{We note that the same designation of frequencies ($f_1,f_2,f_3$) was used for both stars. Symbols used in Sect.\,\ref{ogle-blap-001} refer to Table \ref{tab:ob1-final}, those used in Sect.\,\ref{zgp-blap-08}, to Table \ref{tab:zb8-combined}.}. The terms with frequencies $f_2$ and $f_3$ were detected already in the preliminary analysis, but only in the ZTF and TESS data and with lower S/N than in the combined data. Harmonics of $f_1$ up to 6$f_1$ were detected in the combined data set, although 5$f_1$ and 6$f_1$ marginally, with S/N below 5.0. We also detected two combination terms, $f_1+f_2$ and $f_1+f_3$. The last two terms listed in Table \ref{tab:zb8-combined}, $f_1^\prime$ and $f_2^\prime$, are likely spurious, as they appear very close to the two strong terms, $f_1$ and $f_2$. They probably occur as a result of combining data taken in different passbands and/or imperfections in the approximation of period changes. We also cannot exclude the possibility that $f_1$ and $f_2$ change in a different way, which can also cause these two frequencies to appear. Figure \ref{fig:zb8-fspectra} shows Fourier frequency spectra of the combined data of ZGP-BLAP-08 for three steps of prewhitening: the original data, after removing $f_1$ and its two strongest harmonics and after removing all 12 terms listed in Table \ref{tab:zb8-combined}. Figure \ref{fig:zb8-phased} shows the phased light curve of ZGP-BLAP-08. The pulsation appears to have a light curve typical of most BLAPs, with a steep increase in light followed by a much slower decrease.

The separations $\Delta_1 = f_2-f_1 = 0.657407(21)$\,d$^{-1}$ and $\Delta_2=f_1-f_3=0.657474(24)$\,d$^{-1}$ differ by 0.000067(32)\,d$^{-1}$, that is, are the same within 2-$\sigma$ uncertainty. In view of the presence of the spurious frequencies, $f_1^\prime$ and $f_2^\prime$, close to the two terms of the triplet, we can safely conclude that the difference is insignificant, and therefore the three independent terms detected in the photometry of ZGP-BLAP-08 form an equidistant triplet, similar to the triplet found in OGLE-BLAP-001 (Sect.\,\ref{ogle-blap-001}).

\section{Discussion and conclusions}\label{sect:discussion}
As we have shown in Sect.\,\ref{targets}, the two BLAPs, OGLE-BLAP-001 and ZGP-BLAP-08, exhibit a very similar picture of variability: a dominant pulsation mode with a non-sinusoidal light curve and two symmetric `side' components equidistant in frequency. The dominant mode is most likely the fundamental radial mode, as shown for OGLE-BLAP-001 by \cite{2017NatAs...1E.166P}. We interpret the other two frequencies as a manifestation of the presence of a magnetic field in these two BLAPs. Our conviction comes from the fact that the side frequencies (denoted $f_2$ and $f_3$ for both stars) and the dominant term ($f_1$) are perfectly equidistant in frequency with a precision of 2.7\,$\times$\,10$^{-5}$ and 1.5\,$\times$\,10$^{-4}$ for OGLE-BLAP-001 and ZGP-BLAP-08, respectively. (These numbers are 3-$\sigma$ errors of the differences in frequency separations divided by the separations.) This precision is afforded by observations that cover $\sim$19 and $\sim$10 years for these two stars, respectively. In addition, the presence of a magnetic field in these BLAPs can be explained in view of the two proposed scenarios for the formation of BLAPs as a product of the merger of two stars. One such scenario was proposed by \cite{2023ApJ...959...24Z}. It assumes that BLAPs are formed by the merger of a main sequence star and a helium white dwarf. In another merger scenario proposed in the accompanying paper of \cite{2024arXiv241000154K}, two low-mass helium white dwarfs merge producing a star, which in some conditions may become a BLAP. OGLE-BLAP-001 and ZGP-BLAP-08 are therefore, in our opinion, magnetic pulsators whose variability can be explained in the oblique pulsator model (Sect.\,\ref{oblique-puls}). Our hypothesis that the two target stars are magnetic can be verified observationally, which we are going to carry out.

In the oblique pulsator model, the frequency differences between the main mode and the `side' modes can be interpreted as the rotation frequency of the star, $|f_1-f_2|=|f_1-f_3|=f_{\rm rot}$.\footnote{In principle, it is also possible that the side frequencies of triplets are separated from the central frequency by 2$f_{\rm rot}$. In this case, the corresponding rotation periods should be doubled and the rotational equatorial velocities divided by 2.} The corresponding rotation periods, $P_{\rm rot}=f_{\rm rot}^{-1}$, are thus equal to 0.525500\,$\pm$\,0.000008\,d for OGLE-BLAP-001 and 1.52106\,$\pm$\,0.00010\,d for ZGP-BLAP-08. These are the first rotation periods known for BLAPs. We estimated the luminosities of both stars from the period-luminosity relation provided by \cite{2024arXiv240416089P}, getting $\log(L/{\rm L}_\odot)=2.15$ for OGLE-BLAP-001 and 2.26 for ZGP-BLAP-08. Then, we adopted effective temperatures, $T_{\rm eff}$, of both stars as 30\,800\,K \citep{2024arXiv240416089P} for OGLE-BLAP-001 and 29\,000\,K for ZGP-BLAP-08. The latter is an average value for BLAPs, as no measurement of $T_{\rm eff}$ is available for this star. This allowed us to estimate the radii of both stars for 0.53\,R$_\odot$ for OGLE-BLAP-001 and 0.42\,R$_\odot$ for ZGP-BLAP-08. These two values translate into equatorial rotational velocities, $V_{\rm eq}=2\pi R/P_{\rm rot}$, which is equal to 51\,{\kms} for OGLE-BLAP-001 and 14\,{\kms} for ZGP-BLAP-08, where $R$ denotes stellar radius. Although for both stars spectra have been taken, by \cite{2017NatAs...1E.166P} and \cite{2024arXiv240416089P} for OGLE-BLAP-001 and by \cite{2022MNRAS.511.4971M} for ZGP-BLAP-08, the authors of these papers did not determine the projected rotational velocities, which could be compared with the $V_{\rm eq}$ values, estimated above. Very likely, the low resolution and low S/N of the spectra did not allow for this.

\begin{figure*}
\centering
\includegraphics[width=0.9\textwidth]{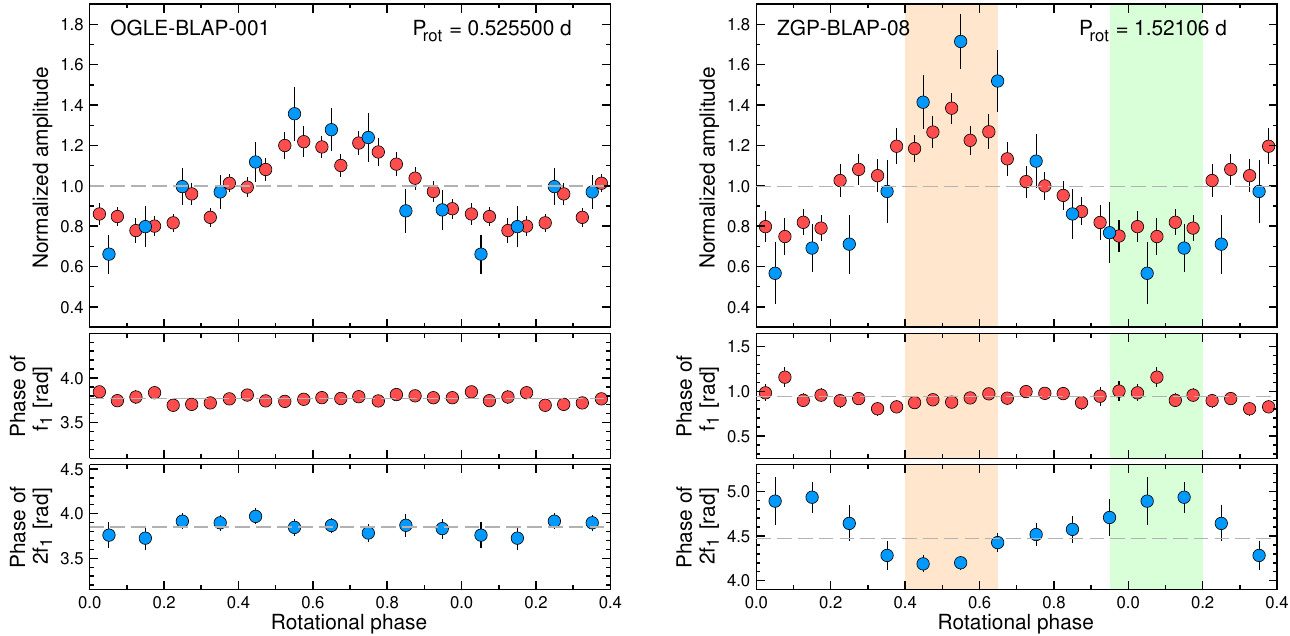}
\caption{Top left: Amplitudes of the $f_1$ term (red) and 2$f_1$ term (blue, normalized by its average amplitude, 0.376, Table \ref{tab:ob1-final}) in the combined data of OGLE-BLAP-001 derived from data in 0.05 ($f_1$) and 0.1 (2$f_1$) phase intervals of the suspected rotational period of 0.525500\,d. Rotational phase 0.0 corresponds to BJD$_{\rm TDB}$\,2457000.0. Middle left: Phases of the $f_1$ term derived from the same data. Bottom left: Phases of the 2$f_1$ term derived from the same data. Top right: Amplitudes of the $f_1$ term (red) and 2$f_1$ term (blue, normalized by its average amplitude, 0.385, Table \ref{tab:zb8-combined}) in the combined data of ZGP-BLAP-08 derived from data in 0.05 ($f_1$) and 0.1 (2$f_1$) phase intervals of the suspected rotational period of 1.52106\,d. Rotational phase 0.0 corresponds to BJD$_{\rm TDB}$\,2459200.0. Middle right: Phases of the $f_1$ term derived from the same data. Bottom right: Phases of the 2$f_1$ term derived from the same data. The orange and light-green strips for ZGP-BLAP-08 panels mark rotational phase intervals near the maximum and the minimum of pulsation amplitude, respectively, for which phased light curves are shown in Fig.\,\ref{fig:zb8-min-max}. The dashed lines in all panels mark the average values of amplitudes and phases from Table \ref{tab:ob1-final} for OGLE-BLAP-001 and from Table \ref{tab:zb8-combined} for ZGP-BLAP-08.}
\label{fig:ap-rot-both}
\end{figure*}

The side frequencies have amplitudes of the order of 10\% of the main term. If our interpretation of the side frequencies in terms of the oblique pulsator model is correct, there is only one (radial) mode excited in each of these two stars, but its visibility is altered by the presence of a magnetic field and should change with the rotational phase. Therefore, we split the combined data for each star into 20 subsets consisting of data in 20 phase intervals of the corresponding $P_{\rm rot}$. The changes of pulsation amplitudes and phases as a function of the rotational phase are shown for both target stars in Fig.\,\ref{fig:ap-rot-both}.

It is clear from this figure that, for both stars, the amplitude of the $f_1$ term changes significantly with the rotational phase. However, the overall behaviour of the light curves for these two stars is different. For OGLE-BLAP-001, the amplitude of the harmonic (2$f_1$) follows the variability of the main term ($f_1$). The harmonic has a slightly higher range of the variability of normalized amplitude, 0.54\,$\pm$\,0.07, than the main term, 0.39\,$\pm$\,0.03. The phases of both terms remain virtually constant. This means that the shape of the pulsation in OGLE-BLAP-001 does not change significantly with the rotational phase.

\begin{figure}[!ht]
\includegraphics[width=\columnwidth]{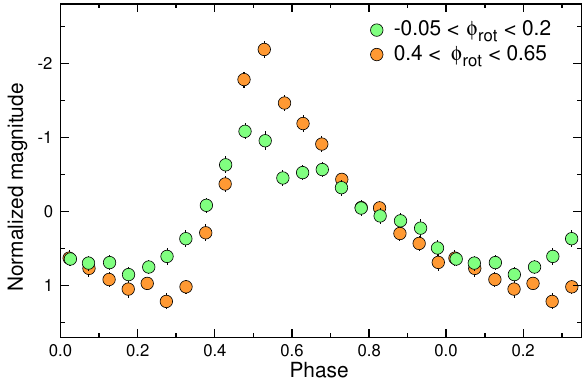}
\caption{Light curve of the combined data of ZGP-BLAP-08 corrected for period changes and phased with the period of $P_1 = 1/f_1 = 0.0244008189$\,d for the two intervals of the rotational phase, $\phi_{\rm rot}$, between $-$0.05 and 0.2 (green dots), that is, near the minimum of amplitude in Fig.\,\ref{fig:ap-rot-both} and between 0.4 and 0.65 (orange dots), that is, near the maximum of amplitude in Fig.\,\ref{fig:ap-rot-both}. The two ranges are marked with similar colours in the right panels of Fig.\,\ref{fig:ap-rot-both}. The magnitudes were calculated in 0.05 phase intervals. Phase 0.0 corresponds to BJD$_{\rm TDB}$ 2459200.0.}
\label{fig:zb8-min-max}
\end{figure}

The dependence on the rotational phase is very different for ZGP-BLAP-08 (right panels of Fig.\,\ref{fig:ap-rot-both}). The range of the variability of the normalized amplitude for the harmonic, 1.01\,$\pm$\,0.10, is twice as large as for the $f_1$ term, 0.52\,$\pm$\,0.03. In addition, while the pulsation phase of $f_1$ remains virtually constant, that of the harmonic shows changes with the full range of about 0.75\,rad. This means that the shape of the light curve of ZGP-BLAP-08 changes with the rotational phase. In order to illustrate this better, we have chosen two intervals of the rotational phase, $\phi_{\rm rot}$, close to the maximum amplitude ($-0.05<\phi_{\rm rot}<0.2$) and close to the minimum amplitude ($0.4<\phi_{\rm rot}<0.65$). The pulsation light curves for these two intervals of the rotational phase are shown in Fig.\,\ref{fig:zb8-min-max}. It can be seen that light curves at these two rotational phase intervals differ both in amplitude and shape.

This picture of variability is very different from what is usually observed in roAp stars, in which strong variability of pulsation phase with rotational phase is observed with a $\pi$\,rad phase reversal predicted by theory for dipole modes \citep[e.g.][]{1982MNRAS.200..807K,2011A&A...536A..73B}. This difference is likely due to the fact that in the case of the two BLAPs, we deal with distorted radial modes, while in roAp stars, the excited modes are usually dipole or quadrupole non-radial modes. Recent observations of some roAp stars, however, show the examples of the suppressed phase changes and different values of the phase differences between central and side components \citep{2016MNRAS.462..876H,2018MNRAS.476..601H,2021MNRAS.506.5629S}. This is usually explained in terms of distorted quadrupole modes and is qualitatively consistent with what is observed in the two BLAPs discussed in the present paper. 

Even without a measurement of a magnetic field, the presence of a multiplet perfectly equidistant in frequency strongly favours the oblique pulsator explanation. One of the alternative explanations would be a rotational splitting of a non-radial mode, but we know -- at least for OGLE-BLAP-001 -- that the observed mode is radial \citep{2017NatAs...1E.166P}, so that it cannot be rotationally split.Although direct mode identification in BLAPs has not been done yet, there are many strong arguments that these stars pulsate in fundamental radial mode. They are the following: (i) BLAPs are single-mode pulsators with large amplitudes. Non-radial pulsators usually show multimode behaviour and have smaller amplitudes due to the averaging over the visible stellar disk. (ii) The only exception is OGLE-BLAP-030, which shows three independent modes \citep{2024arXiv240416089P}, but for this star the period ratio of the two of the three modes also indicates for two radial modes (fundamental and first overtone). (iii) Their light curves resemble other radial pulsators. (iv) The amplitudes of pulsation increase towards shorter wavelengths. In consequence, the variability of colour indices and effective temperatures with pulsation period are observed \citep{2017NatAs...1E.166P,2023NatAs...7..223L}. (v) The relation between light and radial-velocity changes fits that of a radial mode \citep{2023NatAs...7..223L}. (vi) Theoretical calculations favour pulsations in radial fundamental mode as the explanation of pulsations in BLAPs \citep[e.g.][]{2020MNRAS.492..232B}. (vii) BLAPs obey period-luminosity relation \citep{2024arXiv240416089P}.

In order to explain the variability of roAp stars, \cite{1985A&A...151..315M} proposed a spot model, which potentially would be also applicable for the two BLAPs. If there were spots on the surfaces of these BLAPs, however, photometric variability with the rotation periods should be observed, as it is in Ap stars. We did not detect variability with the rotation frequencies in our analysis (Sect.\,\ref{targets}), however. We also verified that our analysis did not affect the detectability of such frequencies. This means that frequency triplets in the two BLAPs cannot be explained by the spot model.

Finally, apparent periodic changes of pulsation period due to the light travel-time effect (LTTE) in a binary system can also result in an equidistant multiplet in the frequency spectrum. In such a case, the splittings derived above would correspond to the orbital frequencies. This explanation for the observed multiplets in the two BLAPs is unlikely, however, because applying the formulae given by \cite{2012MNRAS.422..738S}, we obtained unrealistic mass functions, higher than 280\,M$_\odot$, and huge peak-to-peak radial-velocity amplitudes, larger than 2400\,{\kms}. This leaves us with the oblique pulsator model as the only viable explanation for the frequency triplets observed in the two discussed BLAPs.

While the frequency triplets in the two BLAPs cannot be interpreted in terms of LTTE, we observe significant long-term period changes in both stars (Figs.\,\ref{fig:o-c-o1} and \ref{fig:o-c-zb8}). These changes are not consistent with the assumption of a constant rate of period change. Perhaps they can be interpreted similarly to the changes seen in ZGP-BLAP-01 (TMTS-BLAP-1) \citep{2023NatAs...7..223L}, i.e.~by a combination of evolutionary period changes with a constant rate and periodic changes resulting from LTTE in a wide binary system. However, we think it is too early to advocate this interpretation, as the data have gaps and do not include at least some putative orbital cycles. If indeed these two BLAPs are currently components of binary systems, the orbital periods are on the order of several years, while the evolutionary rates of period change are negative. The latter is consistent with the changes predicted for BLAPs, which are products of mergers. However, these stars certainly deserve long-term photometric monitoring to determine the nature of period changes. 

\begin{acknowledgements} 
We thank the anonymous referee for useful comments and suggestions. This work was supported by the National Science Centre, Poland, grant no.~2022/45/B/ST9/03862. This work used the SIMBAD and Vizier services operated by Centre des Donn\'ees astronomiques de Strasbourg (France), and bibliographic references from the Astrophysics Data System maintained by SAO/NASA.
This work has made use of data from the European Space Agency (ESA) mission {\it Gaia} (\url{https://www.cosmos.esa.int/gaia}), processed by the {\it Gaia} Data Processing and Analysis Consortium (DPAC, \url{https://www.cosmos.esa.int/web/gaia/dpac/consortium}). Funding for the DPAC has been provided by national institutions, in particular the institutions participating in the {\it Gaia} Multilateral Agreement.
This paper includes data collected by the TESS mission. Funding for the TESS mission is provided by the NASA's Science Mission Directorate.
This work used data collected by the Optical Gravitational Lensing Experiment (OGLE) maintained by University of Warsaw, Poland.
This work has made use of data from the Asteroid Terrestrial-impact Last Alert System (ATLAS) project. The Asteroid Terrestrial-impact Last Alert System (ATLAS) project is primarily funded to search for near earth asteroids through NASA grants NN12AR55G, 80NSSC18K0284, and 80NSSC18K1575; byproducts of the NEO search include images and catalogs from the survey area. This work was partially funded by Kepler/K2 grant J1944/80NSSC19K0112 and HST GO-15889, and STFC grants ST/T000198/1 and ST/S006109/1. The ATLAS science products have been made possible through the contributions of the University of Hawaii Institute for Astronomy, the Queen’s University Belfast, the Space Telescope Science Institute, the South African Astronomical Observatory, and The Millennium Institute of Astrophysics (MAS), Chile.
Based on observations obtained with the Samuel Oschin Telescope 48-inch and the 60-inch Telescope at the Palomar Observatory as part of the Zwicky Transient Facility project. ZTF is supported by the National Science Foundation under Grants No.~AST-1440341 and AST-2034437 and a collaboration including current partners Caltech, IPAC, the Oskar Klein Center at Stockholm University, the University of Maryland, University of California, Berkeley, the University of Wisconsin at Milwaukee, University of Warwick, Ruhr University, Cornell University, Northwestern University and Drexel University. Operations are conducted by COO, IPAC, and UW.
\end{acknowledgements}
\bibliography{BLAPs}

\begin{appendix}
\section{Times of maximum light for the target stars}
\begin{table}[!hb]
\caption{Times of the maximum light of the dominant term with frequency of 50.96\,d$^{-1}$ for the analyzed data of OGLE-BLAP-001.}
\label{tab:tmax-ob1}
\centering
\small
\begin{tabular}{r@{.}lrr@{.}ll}
\hline\hline\noalign{\smallskip}
\multicolumn{2}{c}{$T_{\rm max}$} & \multicolumn{1}{c}{$E$} & \multicolumn{2}{c}{$(O-C)$} & Data, season(s) \\
\multicolumn{2}{c}{BJD$_{\rm TDB}$\,$-$\,2450000} &  & \multicolumn{2}{c}{[min]} & filter(s) \\
\noalign{\smallskip}\hline\noalign{\smallskip}
3427&40341(4) &      0 &   0&00(6)  & OGLE-III, 2005, $I$\\
3481&53925(4) &   2759 &   0&28(6)  & OGLE-III, 2005, $I$\\
3768&70044(10) & 17394 &   1&63(14) & OGLE-III, 2006, $I$\\
6442&12620(38) & 153644 &   1&72(53) & OGLE-IV, 2013, $I$\\
6743&72849(20) & 169015 &   2&62(28) & OGLE-IV, 2014, $I$\\
7117&26532(27) & 188052 &   6&72(38) & OGLE-IV, 2015, $I$\\
7474&49528(8)  & 206258 &   8&77(11) & OGLE-IV+Swope,\\
  \multicolumn{2}{c}{}&&\multicolumn{2}{c}{}&2016, $I$ \\
7880&99366(27) & 226975 &   9&24(39) & OGLE-IV,\\
  \multicolumn{2}{c}{}&&\multicolumn{2}{c}{}& 2017+2018, $I$ \\
8576&84890(29) & 262439 &   8&26(42) & OGLE-IV, 2019, $I$\\
8904&58567(13) & 279142 &   7&26(20) & OGLE-IV, 2020, $I$\\
\noalign{\smallskip}\hline\noalign{\smallskip}
  7420&33955(8) & 203498 &  8&10(11) & Swope, 2016, $V$ \\
\noalign{\smallskip}\hline\noalign{\smallskip}  
  9293&44304(50) & 298960 & 5&85(70) & TESS, Sector 36\\
  9319&95107(51) & 300311 & 5&01(73) & TESS, Sector 37\\
 10027&38230(38) & 336365 & 3&06(53) & TESS, Sector 63\\
 10054&49808(27) & 337747 & 1&48(39) & TESS, Sector 64\\
\noalign{\smallskip}\hline\noalign{\smallskip}  
 9639&56535(53) & 316600 &   5&17(74) & ATLAS, 2022, $o$\\
 9683&36045(29) & 318832 & 5&13(41) & ATLAS, 2022, $c$\\
 9730&92295(29) & 321256 &   5&20(42) & ATLAS, 2022, $o$\\
 9955&48877(23) & 332701 &   2&41(32) & ATLAS, 2023, $o$\\
10019&78791(23) & 335978 & 1&79(32) & ATLAS, 2023, $c$\\
10067&54690(27) & 338412 &   2&26(38) & ATLAS, 2023, $o$\\
10334&96570(34) & 352041 &  $-$0&99(48) & ATLAS, 2024, $o$\\
10403&54245(61) & 355536 & $-$1&42(85) & ATLAS, 2024, $c$\\
10422&12375(27) & 356483 &  $-$1&76(38) & ATLAS, 2024, $o$\\
\noalign{\smallskip}\hline\noalign{\smallskip}  
 7162&19731(14) & 190342 &   5&02(21) & {\it Gaia} DR3, $G$\\
 7703&16270(14) & 217912 &   7&01(21) & {\it Gaia} DR3, $G$\\
\noalign{\smallskip}\hline 
\end{tabular}
\end{table}

\begin{table}
\caption{Times of the maximum light of the dominant term with frequency of 40.98\,d$^{-1}$ for the analyzed data of ZGP-BLAP-08.}
\label{tab:tmax-zb8}
\centering
\small
\begin{tabular}{r@{.}lrr@{.}ll}
\hline\hline\noalign{\smallskip}
\multicolumn{2}{c}{$T_{\rm max}$} & \multicolumn{1}{c}{$E$} & \multicolumn{2}{c}{$(O-C)$} & Data, season(s) \\
\multicolumn{2}{c}{BJD$_{\rm TDB}$\,$-$\,2450000} &  & \multicolumn{2}{c}{[min]} & filter(s) \\
\noalign{\smallskip}\hline\noalign{\smallskip}
 8250&62578(35) & $-$1803 & $-$2&70(50) & ZTF, 2018, $g$ \\
 8329&10011(29) &    1413 & $-$0&83(42) & ZTF, 2018, $g$ \\
 8496&66092(37) &    8280 & $-$0&25(53) & ZTF, 2018/19, $g$ \\
 8680&20545(29) &   15802 &    2&03(42) & ZTF, 2019, $g$ \\
 8857&30935(38) &   23060 &    6&02(55) & ZTF, 2019/20, $g$ \\
 9037&14483(28) &   30430 &    8&12(40) & ZTF, 2020, $g$ \\
 9108&56644(34) &   33357 &    8&72(49) & ZTF, 2020, $g$ \\
 9340&47019(78) &   42861 &    6&40(112) & ZTF, 2020/21, $g$ \\
 9471&79316(24) &   48243 &    3&19(35) & ZTF, 2021/22, $g$ \\
 9750&64390(34) &   59671 &    0&60(49) & ZTF, 2022, $g$ \\
 9961&05250(31) &   68294 &    1&11(45) & ZTF, 2022/23, $g$ \\
10195&00777(27) &   77882 &    1&45(39) & ZTF, 2023/24, $g$ \\
\noalign{\smallskip}\hline\noalign{\smallskip}
 8294&62233(27) &       0 &    0&00(39) & ZTF, 2018, $r$ \\
 8337&29952(21) &    1749 &    0&23(30) & ZTF, 2018, $r$ \\
 8351&76890(17) &    2342 & $-$0&21(24) & ZTF, 2018, $r$ \\
 8557&54131(32) &   10775 &    0&25(46) & ZTF, 2018/19, $r$ \\
 8687&94110(35) &   16119 &    2&88(50) & ZTF, 2019, $r$ \\
 8701&82497(38) &   16688 &    2&60(55) & ZTF, 2019, $r$ \\
 8892&73973(30) &   24512 &    6&59(43) & ZTF, 2019/20, $r$ \\
 9110&49431(32) &   33436 &    9&02(46) & ZTF, 2020, $r$ \\
 9273&17222(52) &   40103 &    5&65(75) & ZTF, 2020/21, $r$ \\
 9512&95683(31) &   49930 &    2&46(45) & ZTF, 2021/22, $r$ \\
 9811&62139(32) &   62170 &    0&38(46) & ZTF, 2022/23, $r$ \\
10182&56346(29) &   77372 &    1&61(42) & ZTF, 2023/24, $r$ \\
\noalign{\smallskip}\hline\noalign{\smallskip}
 9011&35264(84) &   29373 &    7&36(121) & ZTF, 2020, $i$ \\
 9104&56539(62) &   33193 &    9&71(89) & ZTF, 2020, $i$ \\
\noalign{\smallskip}\hline\noalign{\smallskip}
 7563&56345(204) &  $-$29960 &$-$14&98(294) & ATLAS, \\
  \multicolumn{2}{c}{}&&\multicolumn{2}{c}{}&2015+2016, $o$ \\
 7931&70606(51) &$-$14873 & $-$4&20(73) & ATLAS, 2017, $o$ \\
 8018&08554(53) &$-$11333 & $-$3&36(76) & ATLAS, 2017, $o$ \\
 8297&52574(46) &     119 & $-$0&41(66) & ATLAS, 2018, $o$ \\
 8385&66160(31) &    3731 & $-$0&26(45) & ATLAS, 2018, $o$ \\
 8639&28497(55) &   14125 &    1&58(79) & ATLAS, 2019, $o$ \\
 8740&84240(53) &   18287 &    3&36(76) & ATLAS, 2019, $o$ \\
 8989&88102(48) &   28493 &    8&94(69) & ATLAS, 2020, $o$ \\
 9108&17523(35) &   33341 &    7&57(50) & ATLAS, 2020, $o$ \\
 9417&74472(61) &   46028 &    2&28(88) & ATLAS, 2021, $o$ \\
 9731&12376(40) &   58871 &    1&34(58) & ATLAS, 2022, $o$ \\
 9828&89667(32) &   62878 & $-$0&33(46) & ATLAS, 2022, $o$ \\
10070&39203(48) &   72775 &    0&35(69) & ATLAS, 2023, $o$ \\
10206&76869(52) &   78364 &    1&06(75) & ATLAS, 2023, $o$ \\
10422&32411(37) &   87198 & $-$0&95(53) & ATLAS, 2024, $o$ \\
\noalign{\smallskip}\hline\noalign{\smallskip}
 7299&18483(184)&$-$40795 & $-$8&88(265) & ATLAS, 2015, $c$ \\
 7626&88572(112)&$-$27365 &$-$11&88(161) & ATLAS, 2016, $c$ \\
 8021&57453(92) &$-$11190 & $-$3&83(132) & ATLAS, 2017, $c$ \\
 8365&13996(55) &    2890 & $-$1&05(79) & ATLAS, 2018, $c$ \\
 8697&62720(79) &   16516 &    1&41(114) & ATLAS, 2019, $c$ \\
 9043&85464(52) &   30705 &    7&52(75) & ATLAS, 2020, $c$ \\
 9394&17420(64) &   45062 &    3&25(92) & ATLAS, 2021, $c$ \\
 9801&34826(44) &   61749 & $-$0&17(63) & ATLAS, 2022, $c$ \\
10183&12527(37) &   77395 &    2&46(53) & ATLAS, \\
  \multicolumn{2}{c}{}&&\multicolumn{2}{c}{}&2023+2024, $c$ \\
\noalign{\smallskip}\hline\noalign{\smallskip}
 8693&94432(32) &   16365 &    3&77(46) & OGLE-IV,\\
  \multicolumn{2}{c}{}&&\multicolumn{2}{c}{}&2019, $I$ \\
\noalign{\smallskip}\hline\noalign{\smallskip}
 9782&63299(17) &   60982 &    0&05(24) & TESS S54, 2022 \\
\noalign{\smallskip}\hline\noalign{\smallskip}
 7035&31101(46) &$-$51609 &$-$13&76(66) & {\it Gaia} DR3, $G$ \\
 7609&19650(77) &$-$28090 & $-$9&91(111) & {\it Gaia} DR3, $G$ \\
\noalign{\smallskip}\hline 
\end{tabular}
\end{table}
\end{appendix}
\end{document}